# SIZE OF NANOOBJECTS IN OIL AND GAS SPECIES AND MATERIALS WITH POSITRON ANNIHILATION SPECTROSCOPY

V. I. Grafutin[1], E. P. Prokopev*[2], L. V. Elnikova[3]

*A. I. Alikhanov Institute for Theoretical and Experimental Physics (ITEP), B. Cheremushkinskaya street, 25, Moscow117218, Russian Federation, tel:+7(499)123-63-97 fax:+7(499)127-08-33*

e-mails: [1]grafutin@itep.ru, [2]eugeny.prokopjev@yandex.ru, epprokopiev@mail.ru, [3]elnikova@itep.ru

**ABSTRACT:** The analytical method to determine geometry and size of nano-scale defects in oil and gas species and materials is proposed. The modeling is carried out with the parameters of the positron spectra in the angular distribution method of positron annihilation spectroscopy, and is based on the 'free electron' approximation. From the annihilation decay kinetics, it is possible to express the trapping velocity of parapositronium in pores via intensities of the positronium components and to define the concentration and radii of pores in a porous layer. As the result, size and the concentration of micro-porous cylindrical nano-objects in the silicon samples are estimated.

It is known [1] that positrons are effective probes for the free volume of nano-objects of angstrom and nanometer sizes in metals, alloys, semiconductors, and porous systems. In the Projects for research of oil and gas species and materials (http://www.portalus.ru/modules/science/data/files/prokopiev/Prokopev-Pos-OilPor.pdf), the determination of size of nano-objects is very important. The Project requires comprehensive studies of the defect structure of oil and gas species and materials containing nanometer-sized cavities (vacancies, vacancy clusters, pores) with the different methods of positron annihilation spectroscopy (PAS). It connects the experimentally measured parameters of the annihilation spectra and characteristics of nano-defects (type, size, concentration) in these materials. The behavior of radiating nanometer defects plays a crucial role in industrial materials. This study contributes into the basic knowledge on the porous structure of these materials, the development of theoretical models describing the behavior and properties of nano-scale defects. PAS is one of the most effective methods for the determination the average size of cylindrical and spherical nano-objects (the free volume of pores, cavities, voids), their average concentration and the chemical composition [1-3]. So, PAS is a suitable instrument to study of technologically important nano-materials, including porous systems and some defective materials. PAS allows us to determine the average values $V_{rad} = (4/3)\pi R^3 N$, a percentage of a free space, formed in materials for electronic and nuclear technology in their operations. We discuss the idea to learn correlations between the values $V_{rad}$ and electrical, mechanical and other properties of materials, such as the permeability and the mechanical stress fracture. A brief review of experimental studies of nano-objects in semiconductors and steels of various sorts, which are used as structural materials in modern nuclear reactors [1], and the research of oil and gas species and materials, may confirm these assumptions. The experimental methods for the determination of the strength and brittleness of metals and alloys irradiated with neutrons play a significant role. It is particularly important to investigate critical defects, strongly violating mechanical and radiation properties of materials.

Basing on the model of a moving particle in a plane limited by a round cylindrical absolutely impenetrable wall [2-4], we got the more correct formulas for the radii $R_c$ of cylindrical (the index 'c') and the spherical (symbol 'sp') nano-pores in units of the width of a component of an angular distribution of annihilation photons (ADAP) $\theta_{1/2}$, and the energies $E_{1c}$ and $E_{1sp}$ of the basic para-positronium (p-Ps) state, annihilated in pores of porous silicon, aluminium dioxide, silicon, as well as in metals and alloys irradiated with protons, neutrons and other charged particles:

$$R_c^0[\text{Å}] = \frac{21.1}{\theta_{1/2}[\text{mrad}]}, R_c^0[\text{Å}] = \left(\frac{30.58}{E_{1c}(eV)}\right)^{1/2}, \quad (1)$$

$$R_{sp}^0[\text{Å}] = \frac{16.6}{\theta_{1/2}[\text{mrad}]}, R_{sp}^0[\text{Å}] = \left(\frac{18.85}{E_{1sp}(eV)}\right)^{1/2}. \quad (2)$$

Here $R$ and $\theta_{1/2}$ are expressed in [Å] and [mrad], respectively. We note, that in formulas (1), (2) and further in (5), (6), the numerical quantity are given in [Å], while the value $\theta_{1/2}$ in [mrad] is actually dimensionless (Tab.1).

Tab.1.

Parameters of investigated samples of porous silicon, features of their production and characteristic of the ADAP spectra [1]

| Sample | Sample characteristics | $I_{g2}=S_{g1}/S_{sum}$ | $I_{g1}=S_{g1}/S_{sum}$ | $I_p=S_p/S_{sum}$ | Note |
|---|---|---|---|---|---|
| 164(1) | Monocrystall | | 0.335±0.031 | 0.665±0.035 | |
| PR86 | Pore Si, <111>, SDB-0.03, h=360-370 mc, HF:$C_2H_5OH$=2:1, J=20 mA/cm$^2$ | 0.015± 0.003 | 0.493± 0.052 | 0.492± 0.044 | Porosity,% 45±3 |

In the Table 1, $h$ is the thickness of silicon plates, <111> is their crystallographic orientation, SDB-0.03 is the mark of silicon plates alloyed by B with the specific resistance 0.03 Ω·cm, $Ig$ =

$S_{gi}/S_{sum}$ ($i$=1,2) is the intensity of Gaussian components, and $I_P = S_p/S_{sum}$ is the intensity of parabolic components in the ADAP spectra ($S_{sum}$ is the total area of an experimental ADAP spectrum, and $S_{gi}$ and $S_p$ are areas of Gaussian and parabolic components in a given spectrum, respectively). *J* is the current density.

For the experimental value $\Theta_2 = 0.8$ mrad in porous silicon (Tab.1), one has received the average radius of cylindrical pores $R = 26.4$ Å nm. Their concentration in a porous layer has obtained equal to $\sim 5.6 \cdot 10^{13}$ cm$^{-3}$. The approach of spherical pores gives rise to $R_{sp} = 20.75$ Å nm, and $N_{sp} \sim 1.3 \cdot 10^{14}$ cm$^{-3}$.

A consideration of the kinetic scheme of annihilation disintegrations and the transformation of the positron and Ps states in a porous layer enables to receive a connection between their trapping velocity $k_{tr}$ by pores and the intensity components $I_{si}$ [1]:

$$k_{tr} = I_{g2} \lambda_p \quad \text{s}^{-1}, \tag{3}$$

where $\lambda_p = \lambda = 0.8 \cdot 10^{10}$ s$^{-1}$ is the velocity of the annihilation disintegration of *p*-Ps, the value $I_{g2} = 0.015$ (Tab. 1) in [1] and $k_{tr}$ in the formula (3); we receive the average *p*-Ps trapping velocity by pores $k_{tr} = 1.2 \cdot 10^{8}$ s$^{-1}$. The trapping velocity $k_{tr}$ can be received, in turn, basing on the well-known expression

$$k_{tr} = \sigma v N_{tr} \quad, \text{s}^{-1}. \tag{4}$$

Here $\sigma$ is denoted the average value of the cross-section of the trapping Ps and a positron by pores (defects); $v$ is the velocity of thermal Ps or a positron; $N_{tr}$ is an average concentration of pores or defects in the porous (defective) area of a crystal, sensitive to thermal volumetric Ps and positron states. Thus, from the resulted expressions, it is possible to define the values $k_{tr}, N_{tr}$, if the parameters $\lambda, \sigma(v), v$ and $R_{tr}$, are known. The average thermal velocity of Ps and a positron at a room temperature $T = 293 K$ was estimated with the formula $v = (8k_0 T / \pi m^*_+)^{1/2} = 7.52 \cdot 10^{6}$ cm/s, for a positron $v = 1.05 \cdot 10^{7}$ cm/s, where $k_0$ is the Boltzmann constant, $m^*_+ = 2m_0$ is an effective mass of Ps, $m^*_+ = m_0$ is an effective mass of a positron, $m_0 = 9.1 \cdot 10^{-28}$ g is a mass of a free positron. We assume the trapping cross-section of positrons and Ps by pores is equals to the average value of a geometrical cross-section of pore (defect) $\sigma = \pi R_{tr}^2 = 1.256 \cdot 10^{-13}$ cm$^2$. From the values $R_{tr} = 2 \cdot 10^{-7}$ cm, and $k_{tr}$ and $v$, defined by the formula (3), we define the concentration of pores $N_{tr} = 1.27 \cdot 10^{14}$ cm$^{-3}$, which are the centers of the p-Ps trapping in a porous layer of silicon.

The experiments [1] have shown, that in a volume of pores of silicon, the main part of positrons is annihilated from the positron states of the non-Ps type. We shall consider that such type positron states are the positrons localized in volume of pores in the same way, as Ps atoms. In this case, formulas (1) and (2) will be transformed into

$$R_c[\overset{0}{A}] = \left(\frac{61.1}{E_{1sp}(eV)}\right)^{1/2} , \qquad (5)$$

$$R_{sp}[\overset{0}{A}] = \left(\frac{37.7}{E_{1sp}(eV)}\right)^{1/2} . \qquad (6)$$

The samples of porous silicon of the sizes 10×20×10 mm$^3$, investigated by a method by positron annihilation spectroscopy (PAS), have been cut out from the whole plates of p – type silicon with the <111> orientation. In the experiments, two samples, designated here as 164 (1) (the initial monocrystal sample), PR 86, PR16, PR17 (the sample of porous silicon received with electrochemical processing in solution HF: C$_2$H$_5$OH, at the density of current J=20 mA/cm$^2$) have been chosen. The parameters of studied plates of silicon and the basic characteristics of the ADAP spectra are resulted in Tab.2.

Tab.2

Characteristics of the ADAP spectra of investigated samples of porous silicon of p-type and parameters of cylindrical pores

| Sample № | $I_g = S_g/S_{sum}$ | $I_p = S_p/S_{sum}$ | $k_{tr} \cdot 10^{-9}$, c$^{-1}$ | $R_{tr}$, Å | $N_{tr} \cdot 10^{-15}$, cm$^{-3}$ |
|---|---|---|---|---|---|
| 164(1) | 0.335±0.031 | 0.665±0.035 | | | |
| PR86 | 0.493±0.052 | 0.492±0.044 | 7.21 | 13 | 1.31 |
| PR16 | 0.483 ± 0.045 | 0.517±0.041 | 6.76 | 13 | 1.23 |
| PR17 | 0.511 ± 0.051 | 0.489±0.044 | 8.00 | 13 | 1.55 |

From Tab.2, the difference between intensities of the Gaussian components $I_g$ (oxidized), that is the oxidized by silicon plates, and $I_g$ (oxidized), that is the initial, non-oxidized plate, in the ADAP spectra, can be written down in the next form:

$$\Delta I_g = I_g \text{ (oxidized)} - I_g \text{ (non-oxidized)} = k_{tr}\tau_1 \qquad (7)$$

That is, the average velocity at the trapping positrons by pores is as follows:

$$k_{tr} = \Delta I_g / \tau_1 . \qquad (8)$$

We estimate the value $k_{tr}$ with the formula (8) for the value $\Delta I_g = 0.665 - 0.493 = 0.172$ [1]. Using this value $\Delta I_g$ in the formula (8), we get $k_{tr} = \Delta I_g / \tau_1 = 7.9 \cdot 10^8$ s$^{-1}$; for $\tau_1 = 2.19 \cdot 10^{-10}$ s,

$k_{tr} = 7.9 \cdot 10^8$ s$^{-1}$. At an annihilation region, sizes of pores and the annihilation energy $E$ of positrons on external valence electrons can be also found, using only the ADAP data. Thus, using this energy, it is also possible to find radii of pores with only the ADAP data. For this purpose, we shall result the expression connecting the energy of an annihilated electron-positron pair with $\theta_{1/2}$ ('*FWHM*' means 'full width half-maximum') [1]:

$$E = 6.9 \cdot 10^{-2} (\theta_2)_g^2 \quad . \tag{9}$$

Here $E$ denotes the energy in eV, and $(\theta_2)_g$ is $FWHM$, the full width of the ADAP curve is in [mrad]. So, for samples of silicon the measured $(\theta_2)_g$ is 11.1 mrad, and it corresponds to the average energy of annihilation electron-positron pairs, which equals to $E = 8.5$ eV, and is caused by the average energy of electrons of an external environment of a silicon atom on a pore wall. This energy can be accepted equal to the energy of an electron in an external environment of an isolated atom of silicon. Thus, one should think, that up to annihilation, a positron and Ps in pore are thermal and the measured energy is defined, generally, by the energy of an electron. The tabulated value of the energy for $Si(3p^2 - {}^3P_0)$ at an electronic external environment of silicon is $E(Si) = 8.1517$ eV [4]. Obviously, the consent of these energies, $E$ and $E(Si)$, is quite satisfactory. Thus, positrons are annihilated generally on external valence electrons of silicon atoms of the pore well. It is possible to believe, that the difference in the energies $E - E(Si) = 0.35$ эВ is caused by the contribution of energy of a bond positron, which is being a pore at the energy of annihilated electron-positron pairs. In this case, to define a cylindrical pore's size, the expression (3) is useful,

$$R_c = \left( \frac{61.1}{E - E(Si)} \right)^{1/2} , \tag{10}$$

Thus, at the value $E = E - E(Si) = 0.35$ eV, pores are of 13.2 Å in size.

Further, with the value $R_{tr} = 13.2$ Å, one has defined an average value of a cross-section of the positron trapping by defects $\sigma_T = 5.5 \cdot 10^{-14}$ cm$^2$. For estimations of the average concentration of pores, the values $k_{tr} = 7.9 \cdot 10^8$ s$^{-1}$, $\sigma_T = 5.5 \cdot 10^{-14}$ cm$^2$ and $v = 10^7$ cm/s were accepted. We received the concentration of pores $N_{tr} = k_{tr} / v \cdot \sigma_T = 1.4 \cdot 10^{15}$ cm$^{-3}$.

If we know the general porosity (45 %) [1] and an average volume of a pore, we can estimate the concentration of pores from simple geometrical reasons and, having compared it with the calculated value $N_{tr}$, we can check up the reliability of accepted assumptions. Average sizes of pores $R_{tr} = 3$ nm, defined above with the formula (2), correspond to their average volumes $V_{tr} = \pi R_{tr}^2 \cdot h = 2.8 \cdot 10^{-16}$ cm$^{-3}$, here $h$ is the thickness of a layer of porous silicon. For the case of 'dense packing', the concentration of such pores, being given from the general porosity 0.45, could be equal to $N_{tr}^G \sim 0.45 / (V_{tr} = 2.8 \cdot 10^{-16}) = 1.6 \cdot 10^{15}$ cm$^{-3}$. Divergences of $N_{tr}^G$ from our results for $N_{tr} = 1.4 \cdot 10^{15}$ cm$^{-3}$ at the certain concentration are not so large. Thus, the samples of porous silicon studied by the ADAP method content micro-porous cylindrical nano-objects of about 1 - 3 nm in size and at the concentration $\sim 10^{15}$ cm$^{-3}$.